\documentclass[a4paper,11pt]{article}
\usepackage{jheppub} % for details on the use of the package, please see the JINST-author-manual
\usepackage{lineno}
\usepackage[english]{babel}

\title{\boldmath Numerical Study of Scalar Field Theory on the Fuzzy Onion}

\author{Matej Hrmo, Samuel Kováčik, Juraj Tekel}\affiliation{Department of Theoretical Physics,\\
Faculty of Mathematics, Physics and Informatics, Comenius University, Slovakia}

\emailAdd{matej.hrmo@fmph.uniba.sk}

\abstract{We study the behaviour of the scalar field theory on the fuzzy onion model -- a three-dimensional matrix model consisting of concentric fuzzy spheres of gradually increasing radii. We use a numerical method of Hamiltonian Monte Carlo simulations to study the phase structure of this theory. We identify the field phases, investigate and describe a phenomenon of dynamical phase transitions and attempt to reconstruct phase transition lines. We compare the results with the well-studied phase structure of the fuzzy sphere. Finally, we identify two boundaries on the phase transitions of the theory, a uniform phase boundary and a critical boundary between the disordered and non-uniform phase.}

\begin{document}
\maketitle
\flushbottom

\section{Introduction}\label{sec1}
It is reasonable to expect the spacetime to have some sort of quantum structure on a scale small enough so that we cannot observe it directly with current experiments \cite{AmelinoCamelia:1997gz, AmelinoCamelia:2008qg, Jacob:2007qj, Burns:2023wzv}. An example of this length scale could be the Planck scale \cite{Doplicher:1994tu, Hossenfelder:2012jw}. The idea is that to distinguish two arbitrarily close points, we would need particles with minuscule wavelengths, in turn resulting in extremely high energy of the particle. Up to a point when the particle forms a black hole, preventing us from observing anything. This makes our spatial resolution blurry. This connects to an area of physics that studies quantum spaces, known as the noncommutative geometry approach \cite{Connes:1994yd, Douglas:2001ba, Szabo:2001kg}, where the blurriness of the space is built into the theory -- one introduces a noncommutativity in the coordinates of the space. This is akin to introducing a new uncertainty principle in coordinates, rendering us indeed unable to distinguish between arbitrarily close points. Noncommutativity can be introduced in various ways, and many models of spaces with a quantum structure have been developed over the past years \cite{Hoppe:1982, Snyder:1946qz, Seiberg:1999vs, Madore:1991bw, Grosse:1994ed, Grosse:1995ar, Balachandran:2001dd, Steinacker:2010rh, Kovacik:2023zab}. As the quantum structure is not accessible for direct experimental exploration, one is often left with studying phenomenological consequences or numerical simulations of physics defined on these spaces --- for example, one can study field theories defined on these spaces.

The fuzzy sphere, originally studied by \cite{Madore:1991bw, Hoppe:1982} is a simple model of a spherically symmetric two-dimensional quantum space. Field theories on this space can be expressed in terms of Hermitian matrices; among the simplest is the scalar field theory with a quartic potential. On an ordinary sphere, this theory has two possible phases with either zero or non-zero vacuum expectation value and a specific phase transition between them. In the case of a fuzzy sphere -- as is the case with many matrix models -- there is a third additional phase. The fields are described by matrices, and in addition to two options: all eigenvalues close to $0$ and all eigenvalues close to a single nonzero value, they can be equally split between two different vacua. In terms of the fields, this corresponds to a striped phase, which is related to the UV/IR mixing, a common feature of fuzzy space physics. In \cite{Kovacik:2018thy, GarciaFlores:2009hf, Ydri_2014}, the triple point of this theory was found, and later analytical analysis of the corresponding matrix models led to a value in reasonable agreement \cite{Tekel:2017nzf}. 

Recently, a fuzzy onion model has been proposed \cite{Kovacik:2023zab}. The space in this case is composed of a series of concentric fuzzy spheres of increasing radius, together forming a three-dimensional quantum space with rotational symmetry. The fuzzy sphere provides a well-explored framework for studying fuzzy field theories and the features of noncommutative spaces. This allows the fuzzy onion model to be a well-controlled playground for studying the effects of quantum structure in three dimensions, granting us access to compare with the phenomenology of commutative physics. A feasible way to perform an HMC study of this model has been proposed \cite{Kovacik:2024is}. Here, we conduct this study, focusing on the interactions between different phases across various layers and novel properties and phenomena that the three-dimensional fuzzy field theory may bring. We show that one may identify the three phases of the fuzzy sphere on the fuzzy onion as well --- all layers of the fuzzy onion align to the same phase. However, an interesting phenomenon of dynamical phase transitions is present on the fuzzy onion, where spontaneous changes occur in the phase configuration across the onion layers. This phenomenon makes it a little more challenging to identify phase transition lines for the fuzzy onion phase diagram.

We start by reviewing the fuzzy sphere, the joining of multiple fuzzy spheres into the fuzzy onion model with a radial derivative operator and defining a scalar field theory on this model in Sections \ref{sec:2} and \ref{sec:3}. We then briefly discuss the Hamiltonian Monte Carlo (HMC) method and our numerical setup in Section \ref{sec:4}. In Section \ref{sec:5}, we summarise the results of our analysis, with emphasis on reconstructing the phase diagram of the fuzzy onion model. 

\section{The Fuzzy Sphere}\label{sec:2}
Let us start with a brief review of the fuzzy sphere and the fuzzy onion model. The fuzzy sphere can be viewed as a quantisation of a classical two-sphere where the coordinate functions obey the classical sphere relation
\begin{equation}
    \sum_{i=1}^3 x_i^2=\rho,
\end{equation}
for a fixed constant (radius) $\rho$, but do not commute. Instead, noncommutativity is introduced in a way that matches the $su(2)$ algebra
\begin{equation}
    [x_i,x_j]=i\lambda_N \varepsilon_{ijk} x_k \neq 0.
\end{equation}
This noncommutativity naturally induces a minimal length scale, governed by the parameter $\lambda_N$. This could be taken to be the Planck length or some other appropriate length scale, depending on the given physical problem. The $su(2)$ structure of the noncommutativity allows us to express our coordinate functions in terms of the $su(2)$ generators. Taking 
\begin{equation}
    x_i=\frac{2\rho}{\sqrt{N^2-1}}L_i^{(N)},
\end{equation}
where $N=2j+1$ is the dimension of the spin-$j$ representation of $su(2)$ generated by $L_i$, we recreate the algebra if we consider $\lambda_N=\frac{2\rho}{\sqrt{N^2-1}}$. We keep the representation dimension index $(N)$ present for future use. The classical sphere could then be recovered in the limit $N\rightarrow\infty$, which we will refer to as the commutative limit. Using a finite-dimensional $su(2)$ representation induces a cut-off in allowed angular momenta. On a classical sphere, the basis of functions is formed by the spherical harmonics. On the fuzzy sphere (with an $N$-dimensional representation), this basis is reduced to what is known as the polarisation tensors $Y_{lm}^{(N)}$, $N\times N$ Hermitian matrices that satisfy
\begin{equation}
    [L_i^{(N)}, [L_i^{(N)},Y_{lm}^{(N)}]]=l(l+1)Y_{lm}^{(N)},~~~~[L_3^{(N)}, Y_{lm}^{(N)}]=m Y_{lm}^{(N)},
\end{equation}
 in analogy with the spherical harmonics. Here $\mathcal{K}_L^{(N)}=[L_i^{(N)}, [L_i^{(N)},\cdot]]$ is the matrix version of the Laplace operator and will play the role the kinetic operator. This basis allows us to describe fields on the fuzzy sphere as $N$-dimensional hermitian matrices
\begin{equation}\label{decomp}
    \Phi^{(N)}=\sum_{l=0}^{N-1}\sum_{m=-l}^lc^{(N)}_{lm}Y_{lm}^{(N)},
\end{equation}
allowing us to study, for example, the quartic-interacting scalar field theory defined by the matrix action
\begin{equation}\label{eq:sph_action}
    S\left[\Phi^{(N)}\right]=\frac{4\pi}{N}\text{tr}_N\left(a^{(N)}\Phi^{(N)}\mathcal{K}_L^{(N)}\Phi^{(N)}+b^{(N)}\left(\Phi^{(N)}\right)^2+c^{(N)}\left(\Phi^{(N)}\right)^4\right),
\end{equation}
where the quadratic (mass) and quartic (interaction) term form the potential part of the theory. Here $\frac{4\pi}{N}\text{tr}_N$ is the analogue of a volume integral over the sphere -- it plays the role of a scalar product on the matrices, much like the integral does for smooth functions. For many more details on the construction of matrix spaces see \cite{Steinacker:2024unq}.

The scalar field theory defined by \eqref{eq:sph_action} has been studied numerically, and the phase diagram of the theory has been reconstructed \cite{Kovacik:2018thy,GarciaFlores:2009hf, Martin_2004}, for a nice review see also \cite{Panero:2016wwx}. Three distinct stable phases were found on the fuzzy sphere. The disordered phase (1-cut symmetric phase), where the eigenvalues oscillate around the zero value; the uniformly ordered -- uniform -- phase (1-cut asymmetric phase) where the eigenvalues oscillate around one of the two minima of the classical potential; and the non-uniform phase, sometimes called the striped phase, where the eigenvalues split between the two minima and oscillate around both simultaneously. The striped phase is a common feature of matrix field theories. A nice visualisation of the fuzzy sphere phases can be found in \cite{Kovacik:2018thy}, figure 4. 

\section{The Fuzzy Onion Model}\label{sec:3}
We briefly review the important notions in the fuzzy onion model. The fuzzy onion model is a three-dimensional model of concentric fuzzy spheres. Let us note that the idea of concentric fuzzy spheres appears in the literature, for example \cite{Hammou_2002,Vitale_2013}, and the fuzzy onion model is one unique construction of such space. The radius of the fuzzy sphere constantly increases by a fixed $\lambda$ between every two spheres, and the dimension $N$ of the representation increases with the radius. For the innermost layer $N=1$, for the next layer $N=2$ and so on, up to $N=M$, where $M$ is the number of layers we take. The matrix describing the field then becomes a block-diagonal matrix consisting of the field matrices on each layer:
\begin{equation}\label{eq:fieldmatFO}
      \Psi=
\begin{pmatrix}
\Phi^{(1)} & ~  & ~  & ~  \\
 ~ & \Phi^{(2)} & ~ & ~ \\
 ~ & ~ & \ddots & ~  \\
 ~ & ~  &  & \Phi^{(M)} 
\end{pmatrix}.
\end{equation}
We would like to study the scalar field theory, where the commutative analogue is given by the action
\begin{equation}
    S[\varphi]=\int_\mathcal{D}\text{d}^nx\left(\varphi\Delta\varphi+\frac{1}{2}m^2\varphi^2+\frac{1}{4!}g\varphi^4\right).
\end{equation}
To do this, we need to perform a volume integral over the onion and define the Laplace operator for the kinetic term. Let us first tend to the latter. We already know the angular part of the Laplace operator -- it amounts to taking the Laplace operator on every layer separately; overall, this yields
\begin{equation}
    \mathcal{K}_L\Psi=R^{-2} 
    \begin{pmatrix}
 \mathcal{K}_L^{(1)}\Phi^{(1)} & ~  & ~  & ~  \\
 ~ &  \mathcal{K}_L^{(2)}\Phi^{(2)} & ~ & ~ \\
 ~ & ~ & \ddots & ~  \\
 ~ & ~  &  &  \mathcal{K}_L^{(M)}\Phi^{(M)} 
\end{pmatrix},
\end{equation}
with $R=\text{diag}(\lambda \mathbb{I}_{1\times1},2\lambda \mathbb{I}_{2\times2},\cdots,M\lambda \mathbb{I}_{M\times M})$ being a matrix of radial distance operator -- block-wise the radius of a given layer. 

The next step is defining a radial derivative. This can be tricky, as the matrices describing the field on different layers have different size. Let us have:
\begin{equation*}
    \Phi^{(N)}=\sum_{l=0}^{N-1}\sum_{m=-l}^l c_{lm}^{(N)}Y_{lm}^{(N)}    
\end{equation*}
\begin{equation*}
    \Phi^{(N+1)}=\sum_{l=0}^{N}\sum_{m=-l}^l c_{lm}^{(N+1)}Y_{lm}^{(N+1)}    
\end{equation*}
We define two maps, schematically $\mathcal{U}:(N)\rightarrow (N+1)$ and $\mathcal{D}:(N+1)\rightarrow (N)$ in the following way:
\begin{equation}\label{eq:up}
\begin{split}
    \mathcal{U}\Phi^{(N)}=\sum^{N}_{l=0}\sum^l_{m=-l}c^{(N+1)}_{lm}Y^{(N+1)}_{lm};\begin{cases}
        c^{(N+1)}_{lm}&=c^{(N)} ~\text{for}~l\leq N-1;\\ c^{(N+1)}_{Nm}&=0.
    \end{cases}
\end{split}
\end{equation}
\begin{equation}
    \mathcal{D}\Phi^{(N+1)}=\sum^{N-1}_{l=0}\sum^l_{m=-l}c^{(N)}_{lm}Y^{(N)}_{lm}; ~~~~ c^{(N)}=c^{(N+1)}~\text{for}~l\leq N-1.
\end{equation}
These maps ``resize'' the matrix one size up or down, using their decomposition into respective polarisation tensors. Essentially, we just keep the coefficients of the decomposition and add extra zeroes/remove excessive coefficients if needed. These maps allow us to define the first and second radial derivative, as now we can subtract matrices of different sizes by resizing them first.
\begin{equation}
\partial^{(N)}_r\Phi^{(N)}=\frac{\mathcal{D}\Phi^{(N+1)}-\mathcal{U}\Phi^{(N-1)}}{2\lambda}.
\end{equation}
\begin{equation}
\partial^{2(N)}_r\Phi^{(N)}=\frac{\mathcal{D}\Phi^{(N+1)}-2\Phi^{(N)}+\mathcal{U}\Phi^{(N-1)}}{\lambda^2}.
\end{equation}

Note that the second derivative is defined in analogy with the second symmetric derivative, not simply as the first derivative applied twice. Having these derivatives, we can now define the radial part of the Laplace operator as 
\begin{equation}\label{eq:radlap}
    \mathcal{K}_R\Psi=\partial^2_r\Psi+2R^{-1}\partial_r \Psi.
\end{equation}
Here, the radial derivative operators on $\Psi$ consist of the radial derivatives for each layer, each acting on the corresponding layer. Note that this cannot produce a block-diagonal matrix, since derivatives always involve more than one layer. 

We are now prepared to write the Laplace operator for the fuzzy onion in a simple fashion:
\begin{equation}
    \mathcal{K}=\mathcal{K}_L+\mathcal{K}_R.
\end{equation}

The question of the volume integral in turn proves to be very simple; we sum over the layers with an extra $r^2dr$ type factor. The volume integral then takes the form of $\sum_{N=1}^M \frac{4\pi(\lambda N)^2}{N} \lambda\text{tr}_N \Phi^{(N)}=\text{Tr}(4\pi\lambda^2 R\Psi)$. Now we are prepared to write the defining matrix action for our scalar field theory.
\begin{equation}\label{action}
    S[\Psi]=4\pi\lambda^2 \text{Tr}~R\left(a~\Psi(\mathcal{K}_R+\mathcal{K}_L)\Psi+b~\Psi^2+c~\Psi^4\right).
\end{equation}
Note that $b$ would inherently be negative, analogous to the scalar field mass term.
We will explore this action using numerical methods. To this end, it is useful to set one parameter to $1$ and rescale the others. We choose $a=1$ and rescale $b\to\tilde{b},~c\to\tilde{c}$. Note that the parameters $\tilde{b}$ and $\tilde{c}$ can still scale with $M$ in general. A more careful approach reveals that, to keep the dependence on $M$ equal across all three terms of the action, one needs to rescale the kinetic term by an additional factor of $R^{-2}$. This essentially amounts to keeping a consistent order of $N$ among the terms of the fuzzy sphere action \eqref{eq:sph_action} for every layer separately, i.e. the correct rescaling of every $a^{(N)}$. Note that this notion, while useful for a numerical approach, is quite unnecessary for analytical computations where we have no desire for the parameters to be unscaled. 

\section{Numerical setup}\label{sec:4}
We analyse the action \eqref{action} using numerical simulations based on the Hamiltonian Monte Carlo (HMC) method. The core of this method lies in the Metropolis algorithm. Having defined a matrix action, we are now capable of computing mean values of observables of the model, defined as:
\begin{equation}
    \left\langle\mathcal{O}(\Psi)\right\rangle=\frac{\int\text{d}\Psi\mathcal{O}(\Psi)\text{e}^{-S(\Psi)}}{\int\text{d}\Psi \text{e}^{-S(\Psi)}},
\end{equation}
with $\text{e}^{-S(\Psi)}$ being the probability measure for the field matrices of type \eqref{eq:fieldmatFO}. The Metropolis algorithm or related methods such as HMC allows us to generate a large set of matrices that satisfy this probability distribution. A little more on this method can be found in Appendix \ref{sec:appendix}.
 
In every simulation, we first perform $10^6$ HMC steps for thermalisation. This allows the simulation to reach an arbitrary point from the initial configuration we choose. We then perform $10^6$ more HMC steps to obtain a matrix dataset, measuring observables every $10^3$ steps. We measure the eigenvalues of the field matrix and the value of the action, as these we will need in further evaluation of the simulations.

An advantage of this method lies in its straightforwardness and robustness as it can be utilised to treat a wide range of actions defining various problems on the fuzzy onion (and other models as well). The main disadvantage is its time consumption. At every Metropolis step, we need to calculate the action value for a given matrix. For the kinetic term, this means decomposing the matrix blocks of \eqref{eq:fieldmatFO} back into polarisation tensors, applying the resizing maps to calculate the radial derivatives, and then transitioning back to matrix blocks. The simulation time increases drastically with the number of layers in our fuzzy onion. Therefore, we have decided to work with a 10-layer fuzzy onion, which proves sufficient for observing novel phenomena and initial probing on a personal computer. Final simulations were conducted on a high-precision computing cluster for a 20-layer fuzzy onion, enabling us to run a large number of long simulations in parallel.

\section{Phases and dynamical transitions}\label{sec:5}
The first question a reader familiar with the scalar field theory on the fuzzy sphere would probably pose is --- how does one intend to define the phases on the fuzzy onion? One would expect to find one of the three fuzzy sphere phases on each of the layers of the fuzzy onion. While it is reasonable to expect at least some alignment of phases due to the radial interaction in the model's construction, it is not a priori clear whether it should be the same phase on every layer or whether one should expect a mixture of fuzzy sphere phases across layers. The results of the simulations provide an answer --- they show that it is common for all of the layers to stay in one of the three stable phases observed on the fuzzy sphere. This is very nontrivial and is a result of the interaction between layers introduced by the radial part of the Laplace operator \eqref{eq:radlap}. This makes defining phases somewhat easier --- we can simply identify the three fuzzy sphere-like phases. Let us note that the innermost layers always behave more-or-less chaotically as they do not have enough eigenvalues to be considered obeying the "large $N$" limit that we aim to approach in the fuzzy sphere theory.

\begin{figure}[h]
    \centering
    \includegraphics[width=\linewidth]{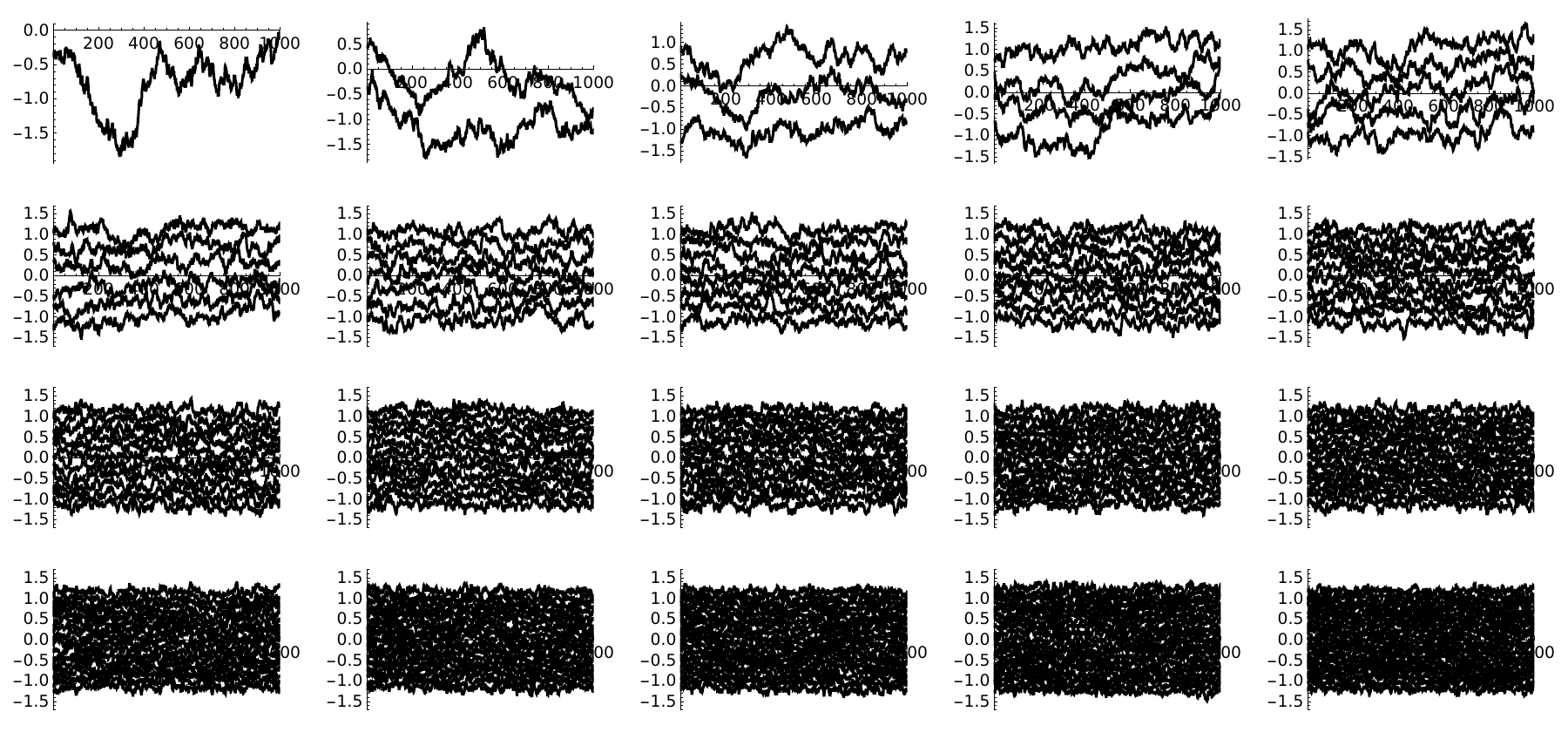}
    \caption{The figure shows the eigenvalue trajectories through the simulation of a 20-layer fuzzy onion for the disordered phase. All of the eigenvalues oscillate around zero. The plots show the various layers, with the layer number growing from left to right and top to bottom. Each plot shows the eigenvalues of that layer as a function of simulation time.}
    \label{fig:disordered20}
\end{figure}

\begin{figure}[h]
    \centering
    \includegraphics[width=\linewidth]{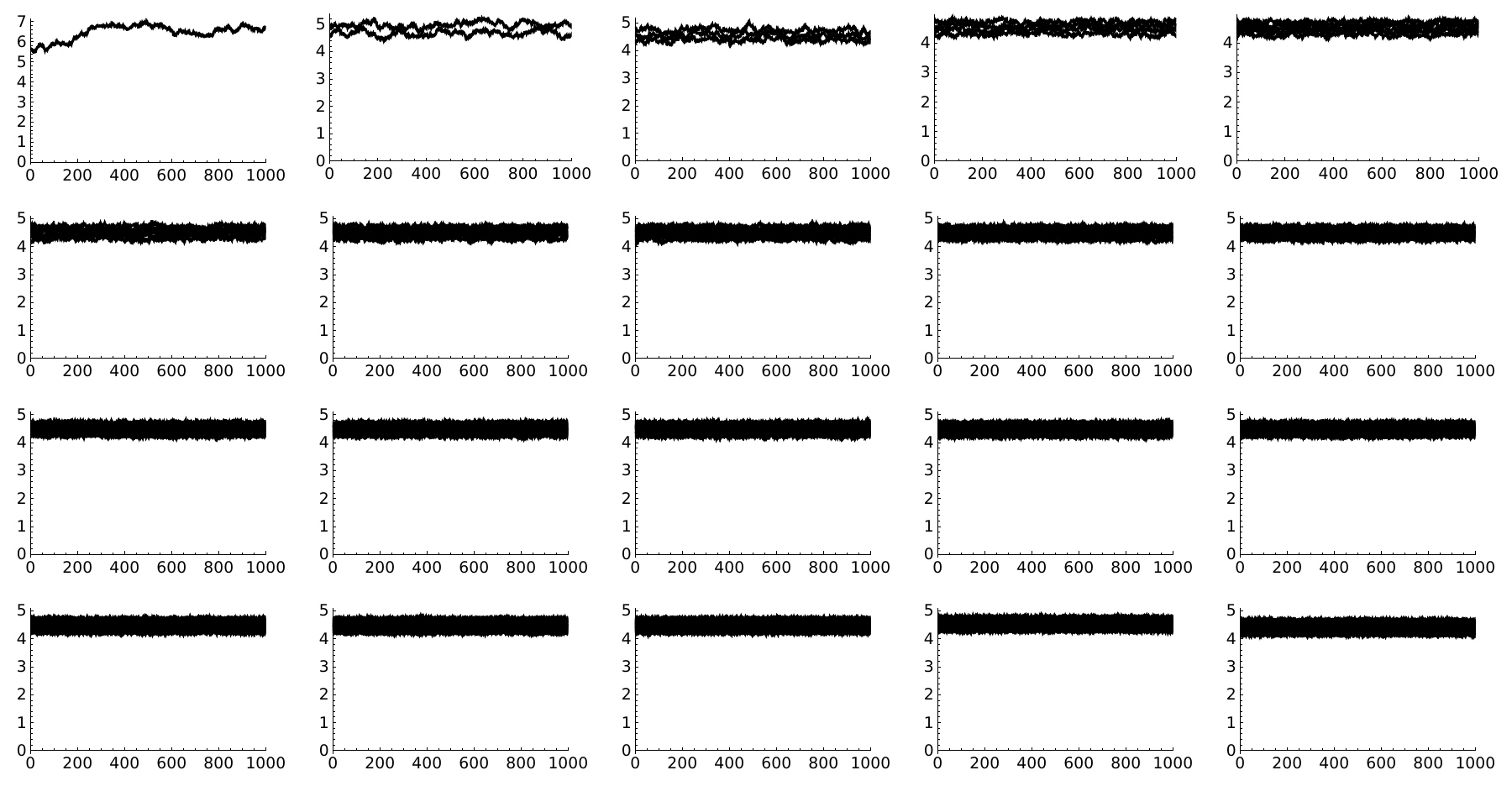}
    \caption{The figure shows the eigenvalue trajectories through the simulation of a 20-layer fuzzy onion for the uniform phase. All of the eigenvalues oscillate around one of the minima (in this case, the positive minimum). The plots show the various layers, with layer numbers increasing from left to right and from top to bottom. Each plot shows the eigenvalues of that layer as a function of simulation time.}
    \label{fig:uniform20}
\end{figure}

\begin{figure}[h]
    \centering
    \includegraphics[width=\linewidth]{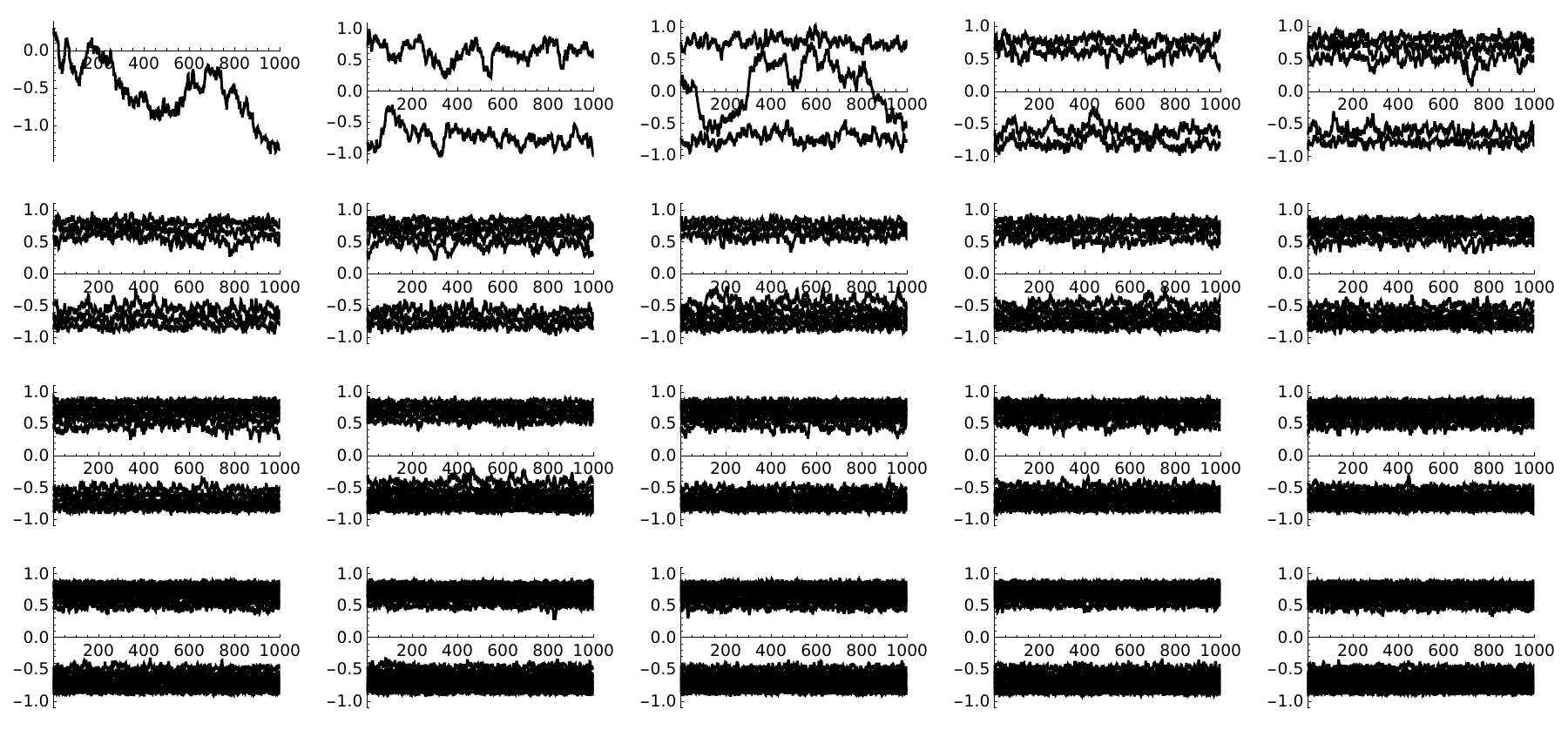}
    \caption{The figure shows the eigenvalue trajectories through the simulation of a 20-layer fuzzy onion for the non-uniform phase. The eigenvalues inhabit both minima of the potential well. The plots show the various layers, with layer numbers increasing from left to right and from top to bottom. Each plot shows the eigenvalues of that layer as a function of simulation time.}
    \label{fig:nonuniform20}
\end{figure}
A novel phenomenon that has already been somewhat described in \cite{Kovacik:2024is} is the dynamical transitions of phase. Even in a well-thermalised simulation, we experience transitions of the eigenvalues between the two minima of the potential well. Often, one of the eigenvalues on the highest layer spontaneously chooses to switch the preferred minimum of the potential. This change then cascades down to the lower layers.

\begin{figure}[h]
    \centering
    \includegraphics[width=0.7\linewidth]{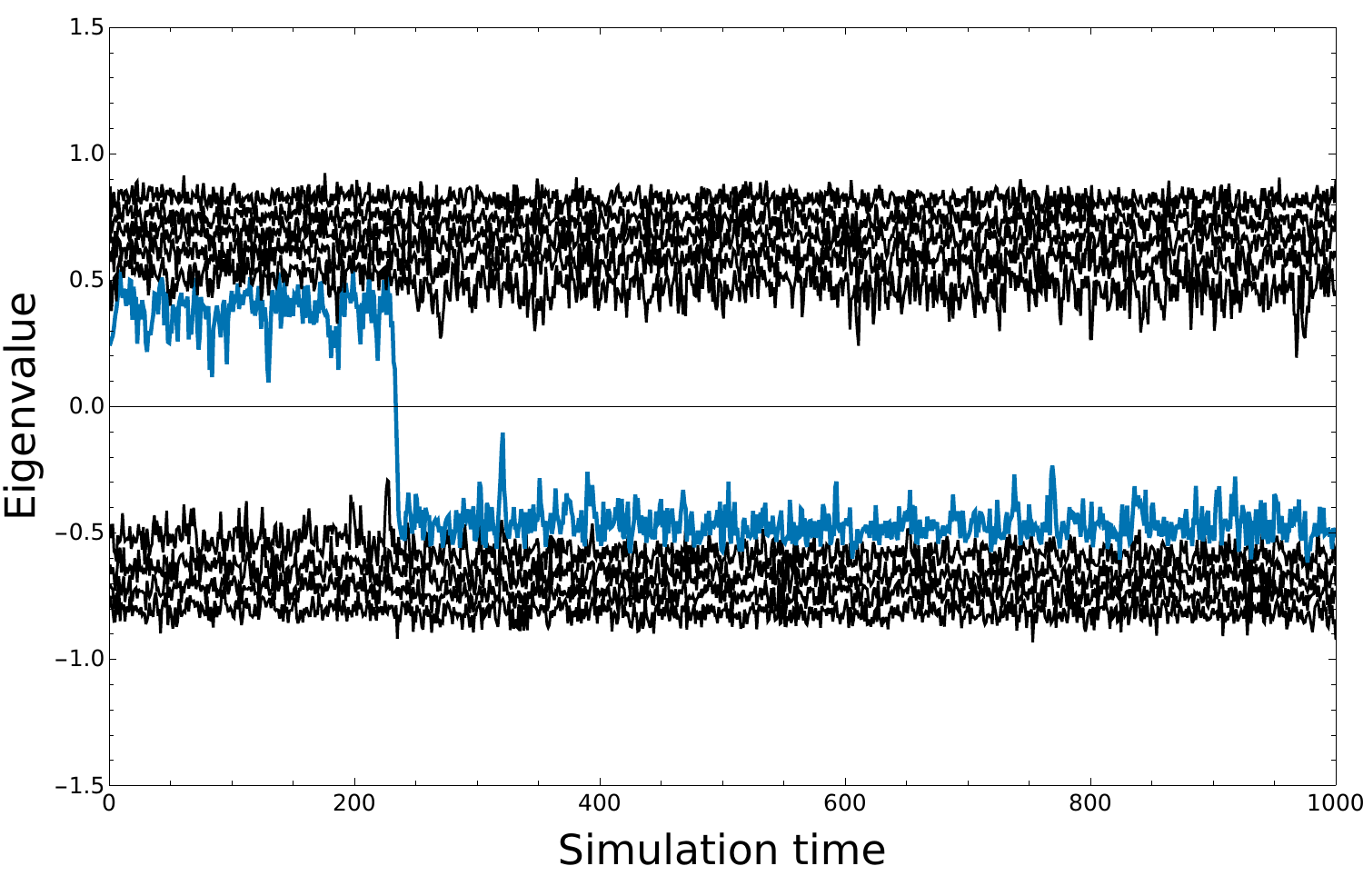}
    \caption{The highlighted eigenvalue spontaneously shifts to the other minimum during the simulation}
    \label{fig:going}
\end{figure}

But this is not always the case. In other cases, sometimes two eigenvalues on the same layer exchange places, if the reader will. One eigenvalue spontaneously changes minima, the other eigenvalue then compensates for the effect; this goes back and forth.
\begin{figure}[h]
    \centering
    \includegraphics[width=0.7\linewidth]{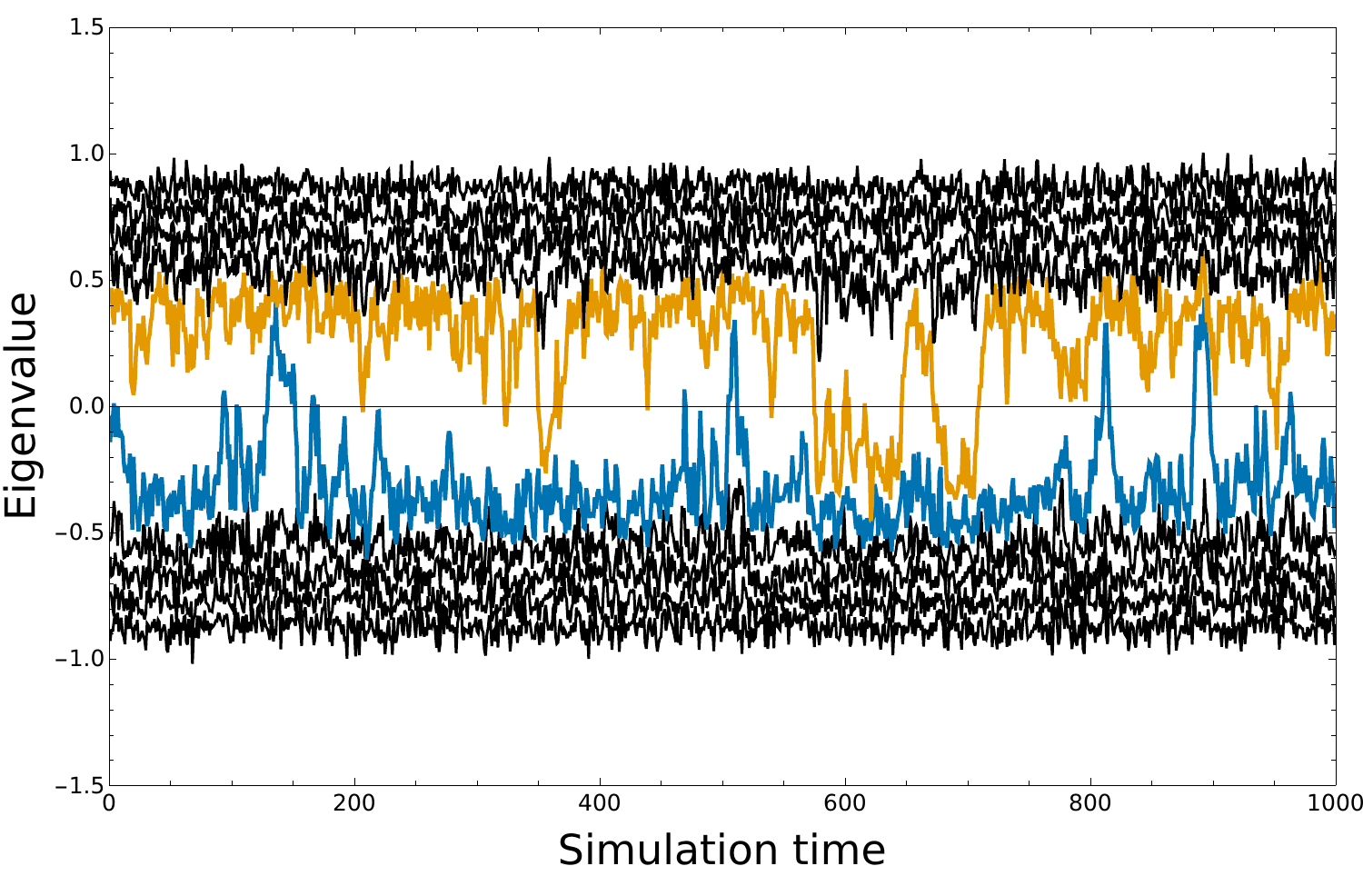}
    \caption{The highlighted eigenvalues spontaneously mix around, spending time in both minima.}
    \label{fig:mixing}
\end{figure}

These are examples of what we call dynamical transitions of phase, an unexpected phenomenon that required further exploration after the initial discovery. The biggest issue is that dynamical transitions blur the phase transitions. Normally, the phase transitions would be sharp lines where the preferred phase (the phase with the lowest free energy) changes, but with the dynamical transitions, the transitions instead become blurred, forming areas in the phase diagram where it becomes very difficult to determine a sharp phase in the fuzzy sphere sense, as described above. These transition areas are present for a 10-layered onion and remain present when we increase the number of layers to 20. Therefore we do not expect this to be a finite-$M$ effect (caused by not enough layers), but instead believe it to be an intrinsic property of the fuzzy onion construction, possibly connected to including fuzzy spheres with a very small dimension of representation (a finite-$N$ effect, if the reader will).

To address these transition areas, we have forced the simulations to start from various initial configurations. We force the simulation to begin in one of the three fuzzy-sphere phases and check whether the phase is stable. This aims to minimise the chance that the dynamical transitions are caused by insufficient thermalisation. 

This has led to two realisations. Firstly, the fuzzy onion simulations are sensitive to the initial configuration. A simulation with identical parameters, initiated in different configurations, ultimately converges to different stable phases. This calls for further examination of the free energies of each solution to determine which is actually preferred, since all obtained solutions appear to be stable. Let us stress that since we generate random Hermitian matrices for the Markov chain in our Monte Carlo and not the eigenvalues of the matrices (despite these being the observable we are interested in), the role of free energy is played by the action $S(\Psi)$ itself, therefore we will consider only that simulation which for the given parameter yields minimal mean action,
\begin{equation}
    \langle S\rangle=\frac{\int \text{d}\Psi S(\Psi)\text{e}^{-S(\Psi)}}{\int\text{d}\Psi\text{e}^{-S(\Psi)}}.
\end{equation}

Secondly, the dynamical transitions did not disappear in any of the initial configurations. We have attempted to solve this by using the result configuration of one simulation as the initial configuration for a simulation with adjacent parameters, but even this still reproduces dynamical transitions. This forces us to admit that blurred phase areas are an apparent feature of the fuzzy onion numerical simulations. However, we are still able to provide some bounds on the phase diagram.

\subsection{Uniformly ordered phase}
We can define the uniformly ordered section of the phase diagram very well. If all eigenvalues have the same sign (no eigenvalue shifts to the other minimum), we call this the uniform phase. This phase is clearly separated from blurred shift-transition area and from the non-uniform phase.

\begin{figure}[h]
    \centering
    \includegraphics[width=\linewidth]{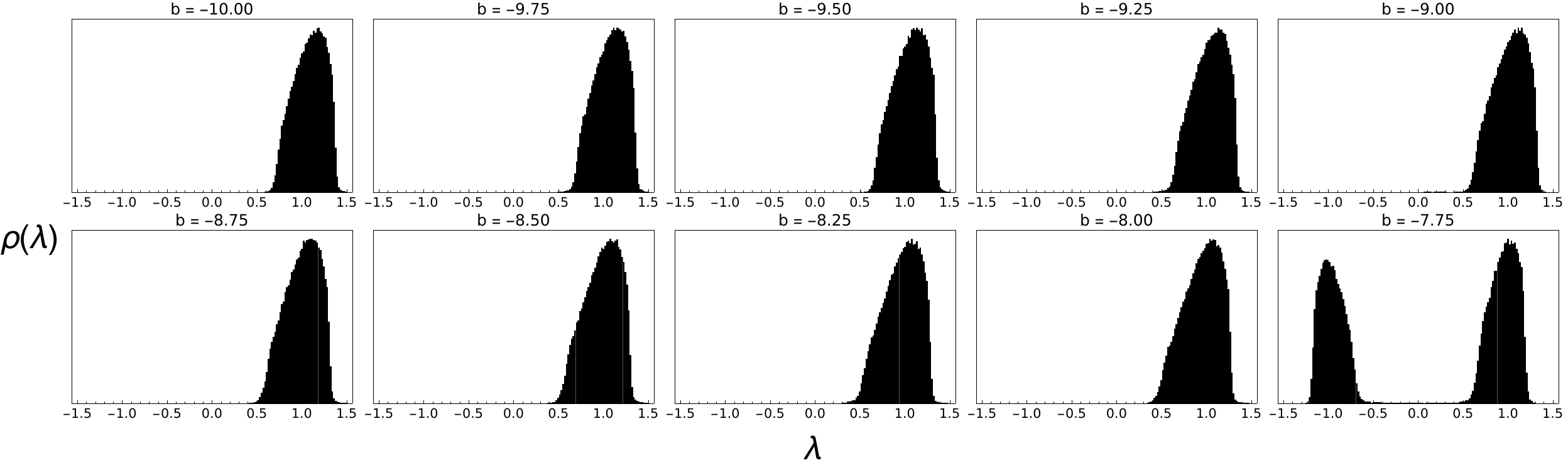}
    \caption{The figure shows the eigenvalue distribution $\rho(\lambda)$ for a fixed $c=4$ and various values of $b$, for a 20-layered onion. The single-peaked uniform phase changes sharply to a two-peaked solution as $|b|$ decreases.}
    \label{fig:uniform_hists}
\end{figure}

\begin{figure}[h]
    \centering
    \includegraphics[width=\linewidth]{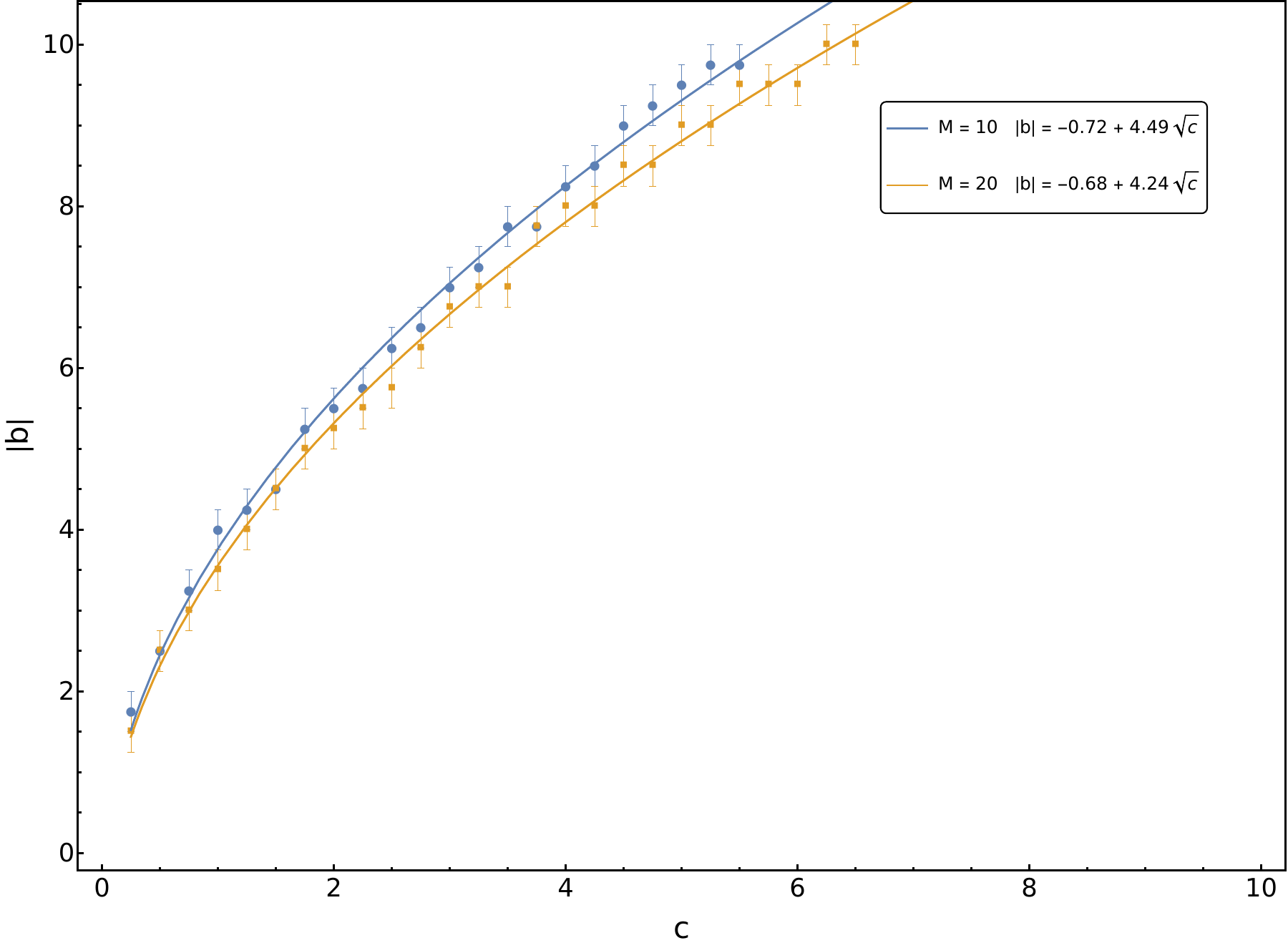}
    \caption{The figure shows the uniform phase boundary obtained from the eigenvalue distributions. For each value of c, the boundary value of $b$ is identified as the largest $\lvert b\rvert$ for which at least s95~\% of the eigenvalues have the same sign --- this is to account for numerical instabilities. The error bars capture this systematic choice of possibly taking a neighbouring datapoint into account. Solid curves show least-squares fits of the form $\lvert b\rvert=k_1+k_2\sqrt{c}$, providing a simple empirical parametrization of the phase boundary. }
    \label{fig:uniform_fit}
\end{figure}

\subsection{Central support extrapolation}
One of the problems with the non-uniform phase definition here, is that for our numerical simulations we almost never see an eigenvalue distribution with two completely separated peaks (a 2-cut solution, if you will).

\begin{figure}[h]
    \centering
    \includegraphics[width=\linewidth]{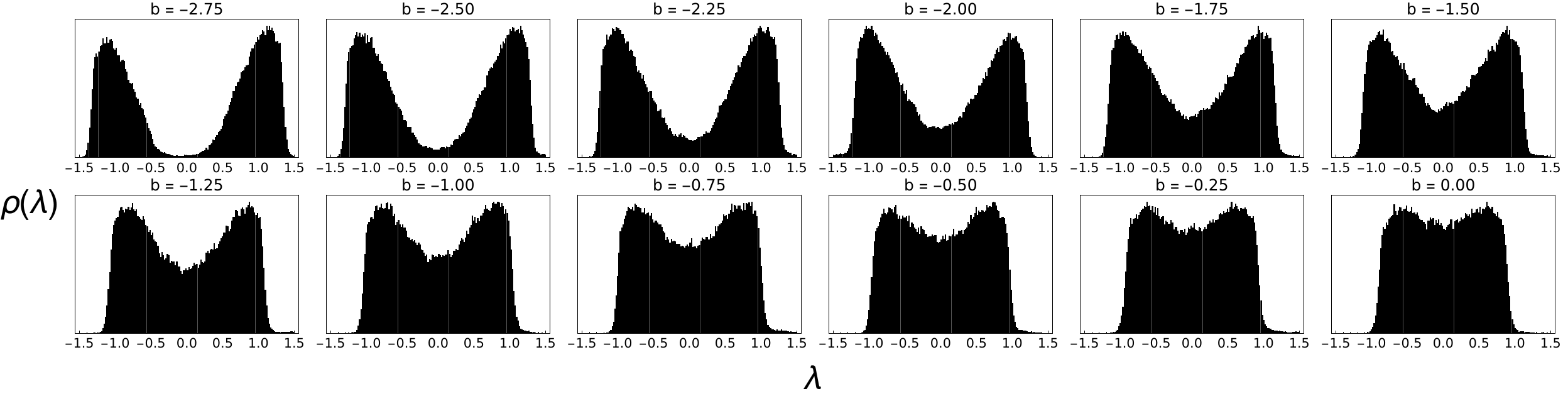}
    \caption{The figure shows the eigenvalue distribution $\rho(\lambda)$ for a fixed $c=1.25$ and various values of $b$, for a 20-layered onion. The central support grows linearly as $|b|$ decreases.}
    \label{fig:rho0_hists}
\end{figure}

\begin{figure}[h]
    \centering
    \includegraphics[width=\linewidth]{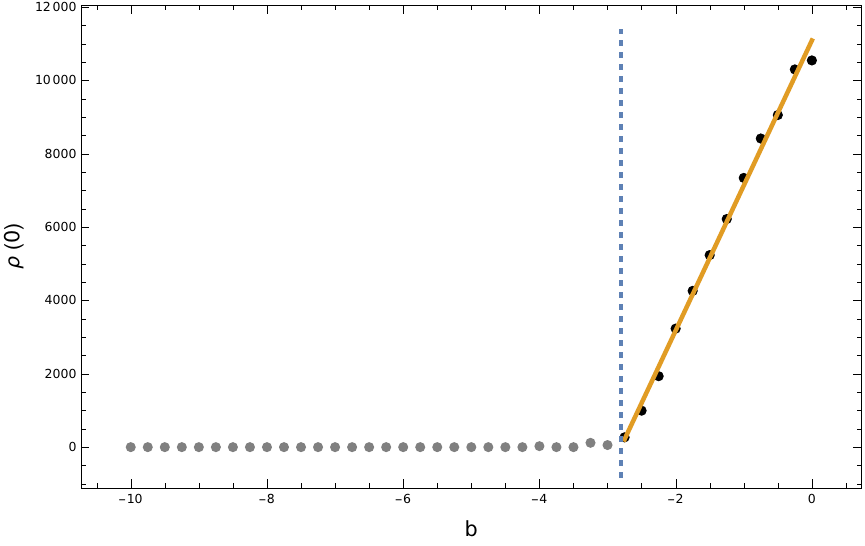}
    \caption{The figure shows the linear growth in the central support from $b_\text{crit}$ to $0$ for a fixed value $c=1.25$. The eigenvalue distributions for this value of $c$ are depicted in Fig. \ref{fig:rho0_hists}.}
    \label{fig:rho0fit}
\end{figure}

A useful notion here is one of the central support. For a fixed $\varepsilon>0$ we can define the central support $\rho_\varepsilon(0)$ as the number (or ratio) of eigenvalues that are inside the interval $(-\varepsilon,\varepsilon)$. The results of our simulation show that the central support is almost never zero, but for large values of $|b|$ (let us remind the reader that, inherently, $b\le0$), it fluctuates on small values. What is of note, however, is that for a fixed value of $c$, there exists a value of $b$ from which the central support seems to grow linearly as we increase $b$ towards $b=0$. For more details see figures \ref{fig:rho0_hists} and \ref{fig:rho0fit}.

A nice property is that the growth remains linear if we decrease $\varepsilon$ (if we approach a more true value of the distribution in zero, $\rho(0)$). If we fit a linear function through the growth, we can identify a theoretical value $b_\text{crit}$, which in theory would have $\rho(0)=0$, where the two peaks of the eigenvalue distributions should separate but they do not -- we assume this to be due to numerical instabilities and effects of finite matrix size.

This critical curve allows us to place an additional bound on the non-uniform phase -- for values of $|b|>|b_\text{crit}|$, non-uniform behaviour would dominate.

\begin{figure}[h]
    \centering
    \includegraphics[width=\linewidth]{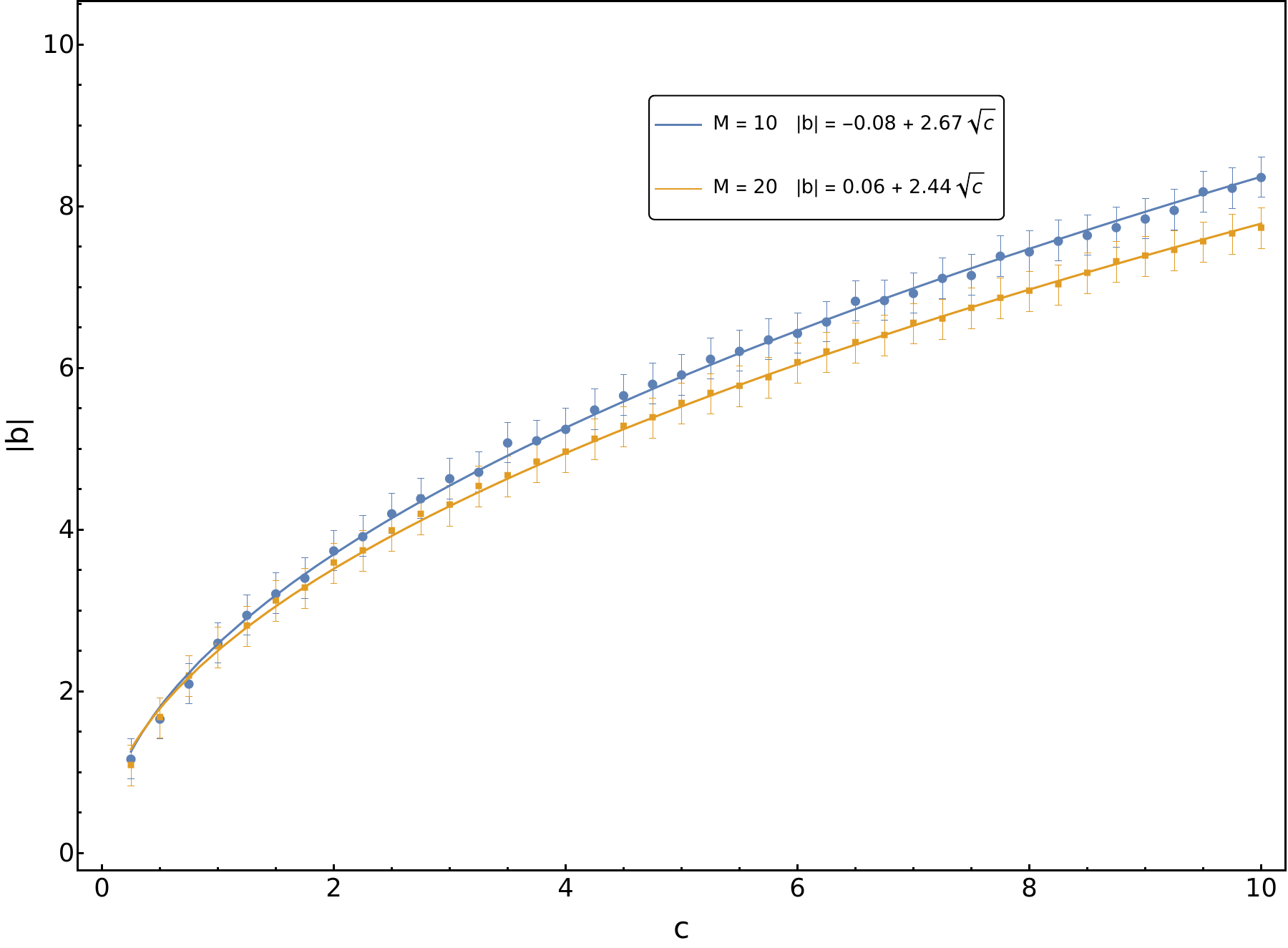}
    \caption{The figure shows the critical curve, where the central support would disappear following the linear dependence. The error bars reflect the possibility of overfitting and underfitting the start of the linear growth. Solid curves show least-squares fits of the form $|b|=k_1+k_2\sqrt{c}$, providing a simple empirical parametrization of the critical boundary.}
    \label{fig:critical_fit}
\end{figure}

\begin{figure}[h]
    \centering
    \includegraphics[width=\linewidth]{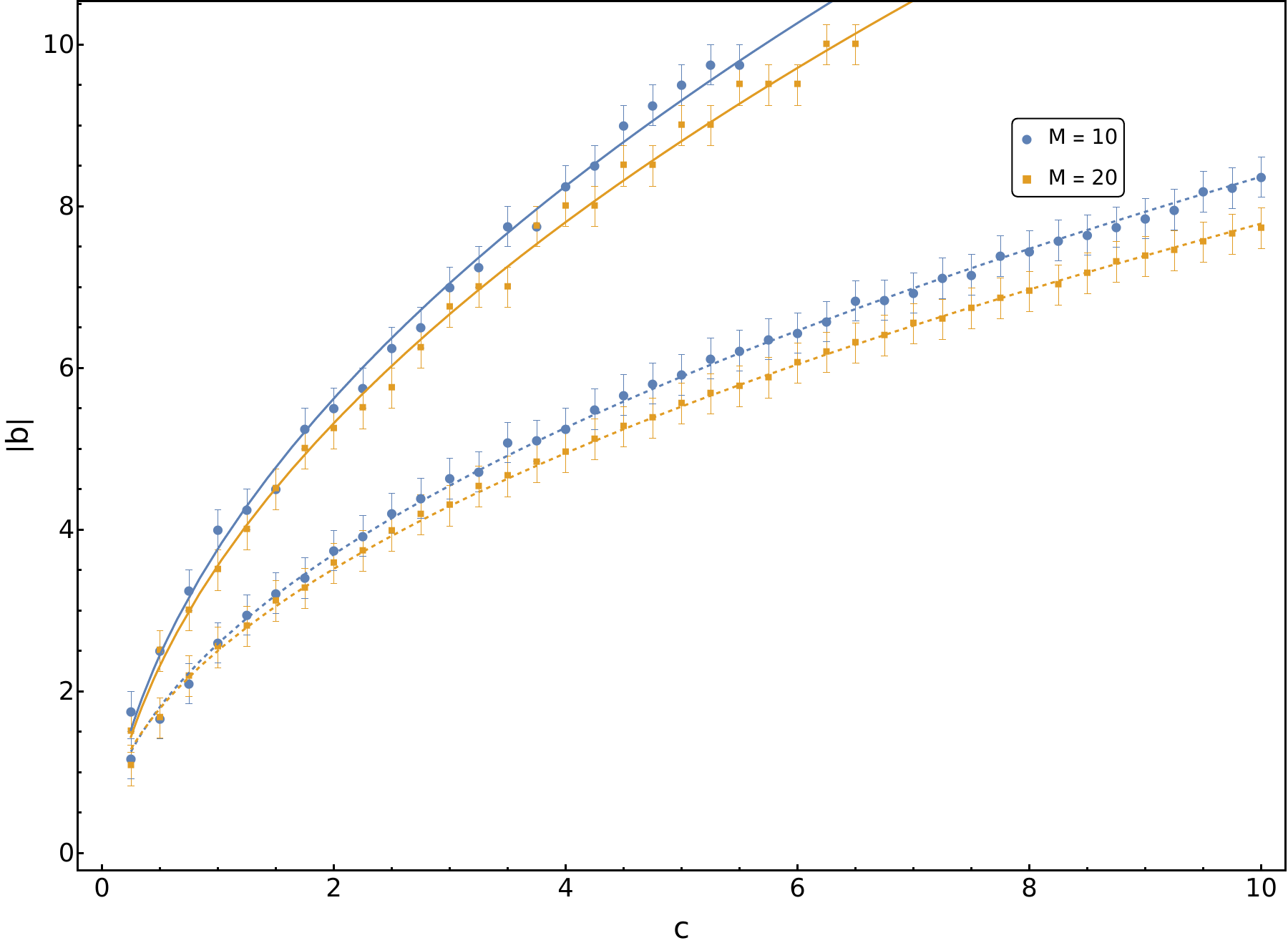}
    \caption{The joint boundaries for a 10-layered and 20-layered onion. The solid line shows the uniform phase boundary. The dashed line shows the critical boundary where the non-uniform phase could be clearly distinguished from the disordered phase in the large $M$ limit.}
    \label{fig:phasediag}
\end{figure}

Altogether, these bounds allow us to split the phase diagram into three areas. One is the clear uniformly ordered phase area; one is the non-uniform blurred area; one is the disordered blurred area. The final phase diagram, and the most important result of the present analysis, is shown in Figure \ref{fig:phasediag}.

\section{Discussion}\label{sec:7}
In this section we discuss the results of our numerical work.
We have found that a scalar field theory on the fuzzy onion model retains a phase structure similar to that of a fuzzy sphere. One can identify a uniform phase, a non-uniform phase and a disordered phase --- consisting of the same fuzzy sphere phase on every layer of the fuzzy onion. Furthermore, the eigenvalue trajectories show new interesting dynamical behaviour. This makes indentifying phase transition lines more difficult than on a single fuzzy sphere. One can get a good handle on separating the uniform phase in terms of same signs of all eigenvalues in the resulting eigenvalue distribution. One can also identify a boundary of critical values $b_\text{crit}(c)$ for which the central support would disappear, marking a distinction between a fully two-cut solution and a single-cut solution, potentially getting some handle on the striped phase. Both of these curves can be well approximated by fits of type $b=k_1+k_2\sqrt{c}$, which differs notably from the fuzzy sphere phase transition lines.

The model does not appear to have a phase transition between the uniform phase and the disordered phase. Let us note that the simulations prove to be very unstable near the origin of the phase space, $|b|,c\sim 0$, therefore it is possible that this transition exists at very small values of parameters.

The fuzzy onion model inherently contains fuzzy spheres with a very small dimension of the representation, therefore the numerical simulations of its field theory suffer from numerical or small-$N$ effects. To suppress these effects, one must work with a very large number of layers, where the number of small-representation layers becomes negligible. Increasing the number of layers $M$, however, carries the difficulty of increasing the field matrix size, which grows as $\frac{M(M+1)}{2}$, on which one must at every Monte Carlo step perform many operations, including complete decomposition into polarisation tensors for the $\mathcal{U}$ and $\mathcal{D}$ operators in the radial derivative. The simulation time therefore rapidly grows as the number of layers increases. The authors hope to continue researching the fuzzy onion field theory with larger onions, using novel high-performance computing clusters.

\appendix
\section{A note on the HMC method}\label{sec:appendix}
Let us briefly comment on the Hamiltonian Monte Carlo (HMC) method as a refresher or for the benefit of a reader not familiar with numerical simulations. At the center of the method is the Metropolis algorithm. The idea is that we start at a random point --- a random matrix --- and then generate a new matrix. If the new matrix has a higher probability than the old one, we accept it and add it to our set. If not, we may still randomly accept it to allow the simulation to escape local minima. Otherwise, we accept the old matrix, save it, and try again, running in a cycle.

This can be further improved by implementing the HMC method itself, which selects the new matrix non-randomly. We instead add a ``momentum'' term $\frac{1}{2}P^2$ to our action, defining $\mathcal{H}(X,P)=S(X)+\frac{1}{2}P^2$ which serves as the Hamilton function. We treat the field matrices as ``position'' matrices here; therefore, we denote them $X$ for better clarity of the analogy. The probability measure becomes $\text{e}^{-\mathcal{H}(X,P)}$. Then, instead of randomly choosing a new matrix in the Metropolis step, we randomly choose a momentum $P$ from a normal distribution (here $P$ is, of course, a matrix). We then let the system evolve according to Hamiltonian dynamics,
\begin{equation}
\begin{split}
    \frac{\text{d}X}{\text{d}t}&=\frac{\text{d}\mathcal{H}}{\text{d}P},\\
    \frac{\text{d}P}{\text{d}t}&=-\frac{\text{d}\mathcal{H}}{\text{d}X},
\end{split}
\end{equation}
arriving at a new field matrix for our Metropolis step. This allows us to achieve a better acceptance rate, as the Hamiltonian, in theory, conserves the Hamiltonian function. In practice, we numerically solve the dynamics, so the solutions of the equations are merely approximations.

\acknowledgments
Part of the research results was obtained using the computational resources procured in the national project National competence centre for high performance computing (project code: 311070AKF2) funded by European Regional Development Fund, EU Structural Funds Informatization of society, Operational Program Integrated Infrastructure.
This research was supported by VEGA 1/0604/26 \emph{Quantum structures of spacetime}. The authors would like to acknowledge the contribution of the COST Action CA23130, \textit{Bridging high and low energies in search of quantum gravity} and COST Action CA21109, \textit{Cartan geometry, Lie, Integrable Systems, quantum group Theories for Applications}. The authors would like to thank Patrik Rusnák for his valuable comments.

%\paragraph{Note added.} This is also a good position for notes added
%after the paper has been written.

% Bibliography

%% [A] Recommended: using JHEP.bst file
\bibliographystyle{JHEP}
\bibliography{biblio.bib}

%% or
%% [B] Manual formatting (see below)
%% (i) We suggest to always provide author, title and journal data or doi:
%% in short all the informations that clearly identify a document.
%% (ii) please avoid comments such as "For a review'', "For some examples",
%% "and references therein" or move them in the text. In general, please leave only references in the bibliography and move all
%% accessory text in footnotes.
%% (iii) Also, please have only one work for each \bibitem.

% \begin{thebibliography}{99}

% \bibitem{a}
% Author,
% \emph{Title},
% \emph{J. Abbrev.} {\bf vol} (year) pg.

% \bibitem{b}
% Author,
% \emph{Title},
% arxiv:1234.5678.

% \bibitem{c}
% Author,
% \emph{Title},
% Publisher (year).

% \end{thebibliography}
\end{document}